\documentclass[reprint,amsmath,amssymb,aps,nofootinbib]{revtex4-1}
\usepackage{float}
\usepackage{xcolor}
\usepackage{graphicx}% Include figure files
\usepackage{dcolumn}% Align table columns on decimal point
\usepackage{bm}% bold math
\begin{document}

\preprint{APS/123-QED}
\title{Electromagnetic Field Enhancement in Bloch Surface Waves}
\author{Daniele Aurelio}
\email{daniele.aurelio01@universitadipavia.it}
\author{Marco Liscidini}
\affiliation{Dipartimento di Fisica, Universit\`{a} degli Studi di Pavia, via Bassi 6, Pavia, Italy}

\date{\today}

\begin{abstract}
We present a systematic comparison between guided modes supported by slab waveguides and Bloch Surface Waves (BSWs) propagating at the surface of truncated periodic multilayers.  We show that, contrary to common belief, the best surface field enhancement achievable for guided modes in a slab waveguide is comparable to that observed for BSWs. At the same time, we demonstrate that, if one is interested in maximizing the electromagnetic energy density at a generic point of a dielectric planar structure, BSWs are often preferable to modes in which light is confined uniquely by total internal reflection. Since these results are wavelength independent and have been obtained by considering a very wide range of refractive indices of the structure constituent materials, we believe they can prove helpful in the design of future structures for the control and the enhancement of the light-matter interaction. 
\end{abstract}

% for a wide range of parameters the smallest mode length in planar dielectric structures is obtained with BSWs rather than  with modes in which light is confined uniquely by total internal reflection. 

\maketitle

%\tableofcontents

\section{Introduction}
When light is confined in one or more dimensions its electromagnetic field is enhanced along with its capacity to interact with matter. In the case of 3D light confinement, this enhancement is usually directly proportional to the square root of the ratio between the dwelling time of light in the resonator and  the mode volume. For instance, this can be exploited to increase and control the spontaneous emission rate of a molecule \cite{purcell1946, Kleppner1981, Yablonovitch1987}. Similarly, in the case of propagating modes, the strength of the light-matter interaction is typically inversely proportional to the square root of the area or length in which light is confined, depending on whether we are dealing with  2D or planar waveguides, and directly proportional to the square root of the time that it spends propagating in the waveguide, which is given by the ratio between the waveguide length and the group velocity. This enhancement is responsible, for example, for the large nonlinear response in silicon nanowires \cite{lipson2006, Foster2008}. Naturally, the size of the mode area (or length) depends on the wavelength under consideration, on the waveguide materials and, above all, on the confinement mechanism, which can be based on total internal reflection (TIR), on interference, or on more exotic phenomena, such as the coupling with free charges in metal, as it happens for surface plasmon polaritons (SPPs) \cite{Kittel_SPP,barnes2006}.

Even considering only the case of planar dielectric structures, one finds a quite surprising variety of confined modes, from D'yakonov waves, which exist at the interface between anisotropic and isotropic media \cite{dyakonov1988, Pulsifer2014}, to guided modes in Bragg waveguides \cite{cho1977, Meade_PBG}. Light confinement in dielectric multilayers has been extensively investigated in fundamental studies of the light-matter interaction and for the development of photonic technologies \cite{Saleh_Photonics,Yeh_OpticalWaves, Yariv_photonics}. Ultimately, the choice of a particular geometry depends on the specific application and on other factors such as material availability and the wavelength range under consideration.  

In the last decade, we witnessed a growing interest in Bloch surface waves (BSWs) \cite{Shinn2005360,History_Meade,liscidini_APL,Descrovi2008, sfez2010,Robertson1,Rodriguez2014, Robertson2}, which are electromagnetic modes that propagate at the interface between a truncated periodic multilayer and a dielectric external medium. Light confinement in BSWs occurs near the multilayer surface and is caused by TIR from the homogeneous layer and by the presence of a photonic band gap (PBG) from the multilayer. Although these modes have been known since the late seventies \cite{History_Yariv, Yeh_78,Robertson19991800,Robertson1993528}, this renewed interest is due to the improvement of those fabrication and growing techniques that today make high-quality multilayers available for a vast class of materials, from semiconductors to oxides and organic compounds \cite{Frezza_2011,Ricciardi_2006}. More in general, other kinds of surface waves have been observed in fully three-dimensional periodic structures, either dielectric or metallic, showing the recent strong interest of the community in asymmetric confinement relying on a PBG on one side \cite{noda2009,tserkezis2011,romanov2016} 

So far, BSWs have primarily been used in all those situations that require the enhancement of the light-matter interaction near the structure surface, with applications ranging from optical sensing to the control of light emission \cite{paeder2011,liscidini_JOSAB,liscidini_2009, Kitada_2017, Munzert_2017, Yu_2014, Descrovi_2010, Kang_2016, Giorgis_2010}. However, despite numerous experimental and theoretical results, it appears as if in many works the authors take for granted that BSWs have a strategic advantage in terms of \emph{surface} field enhancement over simpler solutions such as guided modes in dielectric slab waveguides \cite{khan_2016, Dubey_2017, Angelini2014}. This seems to be in part due to the confusion between the field enhancement, which depends on the area (or length) in which light is confined, and the amount of energy that can be accumulated at the surface of the structure due to the external excitation of the guided mode. 

To clarify this point, let us consider the resonant excitation of a guided mode supported by a generic dielectric planar waveguide. We assume the incoming light to be monochromatic, described by a properly polarized plane wave, and evanescently coupled into the mode through a prism located at a given distance $D$. It is possible to show, by calculating the structure Fresnel coefficients, that the energy accumulated in the guided mode, and thus the electric field at the structure surface,  increases with $D$, i.e. when the coupling strength decreases \cite{sipe_enhancement,Delfan_2012}. Thus, theoretically, it is always possible to achieve any value of the electric field at the surface by simply adjusting the coupling distance $D$ independently of the mode field distribution or the input pump power, making any comparison between different structures in terms of the electric field measured at their surface somewhat arbitrary.

On the contrary, in this communication we present a systematic comparison between dielectric slab waveguides and truncated multilayers in terms of the electromagnetic field enhancement, which is an \emph{intrinsic} property of the guided modes under study. We do this by considering a very wide range of refractive indices for the structure materials to clarify the situations in which BSWs in truncated multilayers are preferable to guided modes in slab waveguides, for example to enhance the interaction with 2D materials or in the design of new types of resonators \cite{Fryett_2016, Chen_2017,Eredi20152076,Menotti_2015, Dubey_2016}. In Sec. II we start by defining the structures under consideration and our figures of merit. In Sec. III we present our numerical results. Finally in Sec. IV we draw our conclusions.

% SECTION 1
\section{Structure geometry and figures of merit}
\begin{figure}
		\centering
		\includegraphics[scale=0.3]{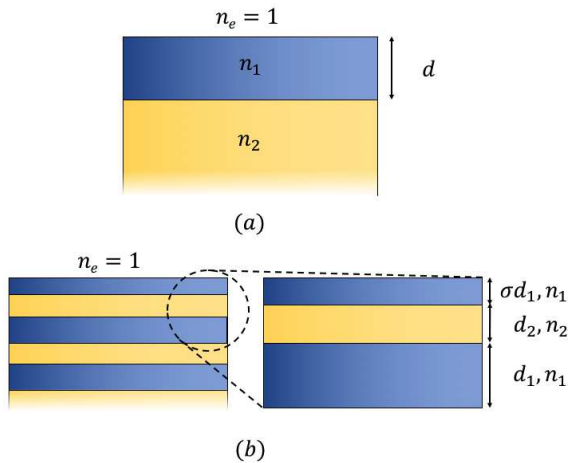}
		\caption{Photonic structure under consideration: (a) asymmetric slab
waveguide of core thickness $d$ and real refractive index $n_{1}=n_{2}+\Delta n$
sandwiched between two semi-infinite dielectric media of refractive
indices $n_{e}$ on the external side and $n_{2}$ on the substrate
side; (b) truncated periodic multilayer with a unit cell consisting of refractive indices $n_{1}=n_{2}+\Delta n$
and $n_{2}$ and of thicknesses $d_{1}$ and $d_{2}$; the structure is semi-infinite and truncated with a
layer of height $\sigma d_{1}$, where $0<\sigma<1$, and refractive index $n_1$.}
		\label{fig1}
	\end{figure}
The structures under consideration are shown in Fig. \ref{fig1} (a) and (b). They are an asymmetric slab waveguide and a truncated periodic multilayer, both in air ($n_e = 1$) and composed of the same two materials having real refractive indices $n_1$ and $n_2$ and refractive index contrast $\Delta n=n_1-n_2>0$. The slab waveguide has a core layer with refractive index $n_1$ and thickness $d$, and a substrate with refractive index $n_2$. In the slab, light confinement occurs uniquely by TIR, with guided modes having effective refractive index $n_2<n_\mathrm{wg}<n_1$ and propagation wave vector $\beta=2\pi n_\mathrm{wg}/\lambda_0$, with $\lambda_0$ the wavelength in vacuum. The periodic multilayer under consideration has a unit cell with two layers having thicknesses $d_1$ and $d_2$ and refractive indices $n_1$ and $n_2$, respectively. The multilayer is terminated with a layer of  thickness $\sigma d_1$ (with $0<\sigma<1$) and refractive index $n_1$. In this structure, light is confined by TIR from the upper cladding and by the PBG from the multilayer. In our analysis we shall consider only confined modes having effective index $1<n_\mathrm{BSW}<n_2$, for which TIR does not occur at any interface within the multilayer. Thus, we focus only on planar dielectric structures in which the multilayer surface is accessible from the upper cladding and, at the same time, we avoid working with membranes, as it would be in the case of a symmetric slab waveguide. This choice is motivated by typical experimental conditions in which one prefers to work with structures having a solid substrate. Finally, we restrict our analysis to the case of modes having the electric field in the plane of the multilayer, i.e. TE (transverse electric)-polarized, for which it is not possible to exploit SPPs to enhance the  field at the structure surface. In Fig. 2 (a) and (b), we show two examples of the electric field distribution for the fundamental TE mode of an asymmetric slab waveguide and the TE BSW supported by a truncated multilayer, respectively. The field distribution has been calculated by using the transfer matrix method \cite{Yariv_photonics}. The results are normalized to the mode wavelength in vacuum to take advantage of the scalability of Maxwell equations \cite{Joanno_book}, with the plots and the results presented in the following sections being scale invariant. 
\begin{figure}
		\centering
		\includegraphics[width=0.51\textwidth]{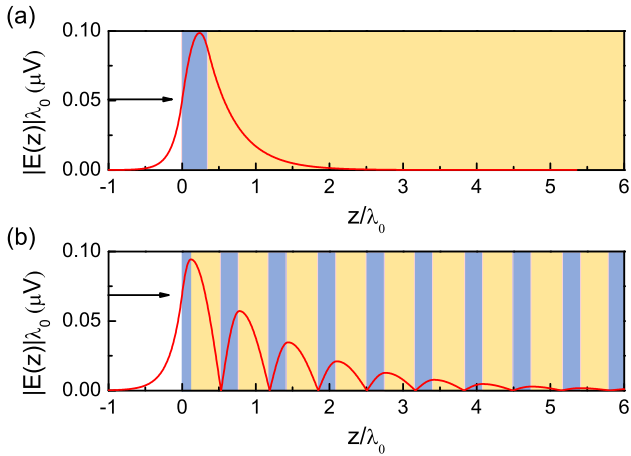}
		\caption{Wavelength-scaled modulus of the electric field of (a) the TE0 mode
supported by an asymmetric slab waveguide of core index $n_{1}=1.7$
and cladding index $n_{2}=1.5$ and $d/\lambda_0 = 0.354$; the core thickness was found  by minimizing the mode length; (b) the BSW supported by a semi-infinite truncated periodic multilayer
with unit cell having indices $n_{1}=1.7$ and $n_{2}=1.5$ and thicknesses $d_1/\lambda_0 = 0.243$, $d_2/\lambda_0 =0.400$; for this structure, $\sigma=0.467$; the structure parameters were found using Eq. \eqref{QW_condition} to minimize the mode length of the BSW.}
		\label{fig2}
\end{figure}

Now, we consider the following questions: (1) Which one of the two structures maximizes the electromagnetic energy density at a generic point along $z$? And (2) which one provides the largest electric field at the structure surface (i.e. $z=0$)? To answer these questions we use the following two figures of merit (FoMs);  the first one is, in strict analogy with 3D resonators \cite{Foresi:1997aa}, the \emph{mode length}, and it can be defined as
\begin{equation}
	L_\mathrm{mod} = \frac{1}{\max \left[\varepsilon(z)|E(z)|^2\right]} \int_{-\infty}^{\infty} \varepsilon(z)|E(z)|^2 dz \label{Lmod},
\end{equation}
where $E(z)$ is the electric field and $\varepsilon(z)=\varepsilon_0n^2(z)$, with $\varepsilon_0$ the vacuum permittivity.  For a fixed amount of energy in the mode, maximizing the electromagnetic energy density in a given point is equivalent to minimizing the mode length $L_\mathrm{mod}$\footnote{It should be noticed that, in spite of its name and units, $L_\mathrm{mod}$ does not measure how tightly light is confined. A measure of the size of the mode can be obtained by calculating 
\begin{equation}
	\sigma^2_\mathrm{mode}=\frac{ \int_{-\infty}^{\infty}(\langle z\rangle-z)^2\varepsilon(z)|E(z)|^2 dz}{\int_{-\infty}^{\infty} \varepsilon(z)|E(z)|^2 dz},
\end{equation} 
with 
\begin{equation}
	\langle z\rangle=\frac{ \int_{-\infty}^{\infty}z\varepsilon(z)|E(z)|^2 dz}{\int_{-\infty}^{\infty} \varepsilon(z)|E(z)|^2 dz}.
\end{equation}}.

The second FoM is simply the value $|E(0)|\lambda_0$ of the modulus of electromagnetic field at the structure surface, taken on the cladding side. In this case, a fair comparison between the structures requires that each mode profile $E(z)$ is properly normalized. Here we use \cite{yang_2008}
\begin{equation}\label{normalization}
	A\int_{-\infty}^{+\infty} \varepsilon(z)|E(z)|^2 dz = \frac{\hbar\omega}{2},
\end{equation}
where $A$ is a normalization area in the plane of the structure (taken to be 1 $m^2$), $\omega=2\pi c/\lambda_0$, and where we have neglected chromatic dispersion of the refractive indices nearby $\lambda_0$. It should be noticed that the choice of normalizing the energy in the mode to that of a photon guarantees that  $L_\mathrm{mod}/\lambda_0$ and  $|E(0)|\lambda_0$ are scale invariant, which leads to energy-independent conclusions. 

% RESULTS AND DISCUSSION
\section{Results and discussion}
Our analysis consists in finding the structures that minimize $L_\mathrm{mod}/\lambda_0$, \emph{i.e.} maximize the field confinement, and/or maximize $|E(0)|\lambda_0$, \emph{i.e.} maximize the field at the surface; this is done for any given pair ($n_1$, $n_2$) of refractive indices, with $n_2\in [1.4,2.1]$ and $n_1\in [1.4,3.4]$. 

In the case of the asymmetric slab waveguide, this task is accomplished by following the semi-analytical approach illustrated in the appendix A:  for any $(n_1,n_2)$, we find the expression of the electric field profile as a function of $d/\lambda_0$, and we search for the values of  $d/\lambda_0$ that maximize the FoMs. The presence of only one independent structure parameter, namely $d/\lambda_0$, makes it easy to determine the two structures that give the best FoMs among \emph{all} the possible asymmetric slab waveguides of the form depicted in Fig.\ref{fig1} (a). 

In the multilayer case, the search for the optimal structure is more challenging, as there are three independent parameters $d_1$, $d_2$, and $\sigma$. To reduce the dimensionality of the problem, we choose $d_1$ and $d_2$ to guarantee the fastest decay of the envelope function of the electric field in the multilayer. This corresponds to the generalized $\lambda/4$ condition:
\begin{equation}\label{QW_condition}
	d_i = \frac{\lambda_0}{4\sqrt{n_i^2 - n^2_{BSW}}},
\end{equation}
assuming a BSW having effective index $n_\mathrm{BSW}$. Under these hypotheses, one considers all the possible effective indices $n_\mathrm{BSW}\in[1,n_2]$ by calculating the corresponding multilayer truncation and searching for the structures that maximize the two FoMs. It should be noticed that, in principle, this approach does not guarantee to find the best structures among all the possible truncated periodic multilayers of the kind shown Fig.1 (b). Yet, our strategy starts from the reasonable assumption that the largest field enhancement is obtained by maximizing the field decay in the multilayer.  In fact, we have verified, by a brute force optimization for some selected ($n_1$, $n_2$) in the range of interest, that the best structures for both the FoMs are either identical to those found through our strategy or do not differ significantly from them. This approach has also the undeniable advantage of providing a quick and handy rule to design the multilayer.  

% FIGURE 3 mode length vs Delta n
\begin{figure}
\includegraphics[width=0.51\textwidth]{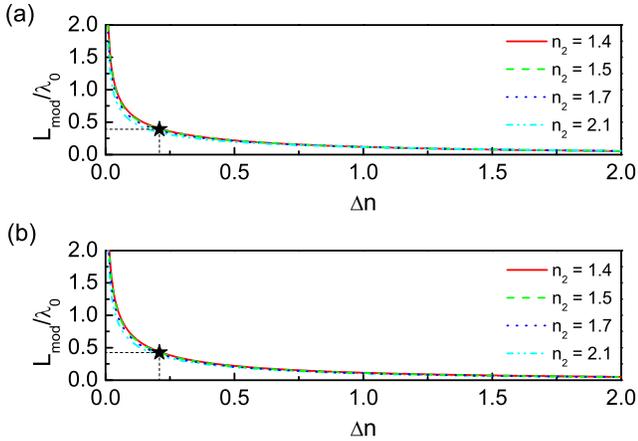}
\caption{(a) Wavelength-scaled mode length $L_{\text{\text{mod }}}/\lambda_0$
of the TE0 mode supported by an asymmetric slab waveguide as a function of the 
index contrast $\Delta n=n_1-n_2$ for selected values of the substrate refractive
index $n_{2}$; (b) scaled mode length $L_{\text{mod }}/\lambda_0$
of the BSW supported by a truncated periodic multilayer as a function of the refractive
index contrast $\Delta n=n_1-n_2$ for selected values of the lower refractive
index $n_{2}$. The stars correspond to the modes shown in fig. 2.}
\label{fig3}
\end{figure}

In Fig.\ref{fig3} we plot the best (i.e. the smallest) $L_\mathrm{mod}/\lambda_0$ for (a) the guided mode in the asymmetric slab waveguide and (b) the BSW in the truncated multilayer as a function of the refractive index contrast $\Delta n=n_1-n_2$. We consider four different values for the low refractive index, namely $n_2=1.4,1.5,1.7$, and $2.1$. There are some important indications that can be obtained by this first set of results: (i) in both structures the mode length is essentially independent of $n_2$, while it is strongly dependent on the refractive index contrast $\Delta n$; (ii) for a given $\Delta n$, the best mode length that can be obtained by TIR in a slab waveguide is similar to that achievable in a truncated multilayer. By looking at Fig. 2 (a) and (b), which correspond to the points indicated in Fig. 3 (a) and (b), one can notice that indeed the maximum of $\varepsilon(z)|E(z)|^2$ in both cases is about the same. This happens in spite of the fact that the BSW field extends in the multilayer deeper than in the case of the asymmetric slab waveguide, showing that $L_{mod}$ is not immediately related to how tightly light is confined. In particular, in the BSW case, the result can be understood as a consequence of the damped \emph{oscillatory behaviour} of the field in the multilayer associated with the presence of the PBG.

% FIGURE 4 mode length ratio

\begin{figure}

		\centering
		\includegraphics[width=0.51\textwidth]{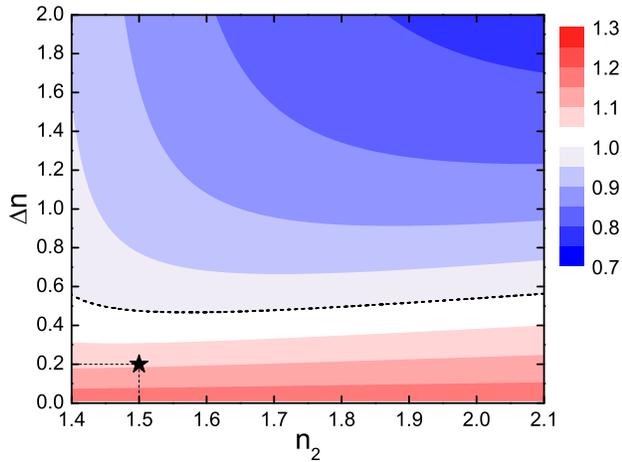}
		\caption{Contour plot of the ratio between the smallest mode lengths of the BSW supported by the truncated periodic multilayer and of the TE0 mode supported by the asymmetric slab waveguide as a function of values of the lower index $n_2$ and the refractive index contrast $\Delta n$. The black dashed line indicates the value for which the modes have the same mode length. The star corresponds to the refractive indices of the structure supporting the modes in fig. 2.}
		\label{fig4}
	\end{figure}	

In Fig.\ref{fig4} we compare the two structures by showing the ratio between the smallest mode length obtainable with the truncated multilayer and that achievable for the asymmetric slab waveguide as a function of $n_2$ and $\Delta n$. The plot confirms the small dependence of $L_\mathrm{mod}/\lambda_0$ on $n_2$ and shows the presence of three different regimes depending on the refractive index contrast:  (i) for $\Delta n> 0.6$ the largest electromagnetic energy density in a point is obtained with a truncated multilayer, and the differences with respect to the slab waveguide mode increases with $\Delta n$; (ii)  for small refractive index contrasts ($\Delta n< 0.5$) the largest electromagnetic energy density in a point is obtained with the asymmetric slab waveguide, and the difference with the truncated multilayer increases as the $\Delta n$ gets smaller; (iii)  There exist an intermediate region, with $0.5<\Delta n< 0.6$, in which the optimization of the two structures leads to the same result. Naturally, these conclusions depend also on our initial choice of having considered air ($n=1$) as our upper cladding, yet a qualitatively similar results is expected also for larger values of the cladding refractive index. Finally, we remind the reader that the results shown in Fig. \ref{fig4} are wavelength independent, because of the normalization condition (\ref{normalization}).
% FIGURE 5 Surface field vs Delta n

	\begin{figure}
		\centering
		\includegraphics[width=0.51\textwidth]{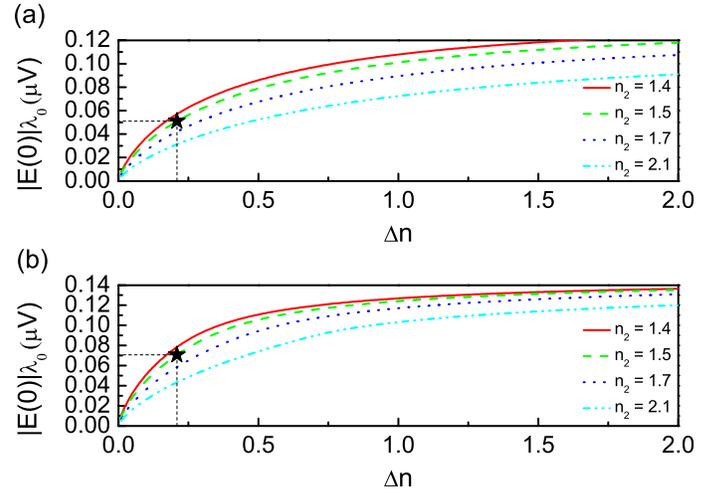}
		\caption{(a) Wavelength-scaled surface field $|E(0)|\lambda_0$ of the TE0 mode supported by an asymmetric slab waveguide as a function of the index contrast $\Delta n=n_1-n_2$ for selected values of the substrate refractive index $n_{2}$. (b) Wavelength-scaled surface field $|E(0)|\lambda$ of the BSW supported by a truncated periodic multilayer as a function of the refractive index contrast $\Delta n=n_1-n_2$ for selected values of the lower refractive index $n_{2}$. The mode field profiles are normalized according to Eq. \eqref{normalization}, and the stars correspond to the modes shown in fig. 2.}
		\label{fig5}
	\end{figure}

We now turn to the analysis of the surface field. In Fig. \ref{fig5} (a) and (b) we show the highest value of the surface field achievable for a guided mode in the asymmetric slab and the truncated multilayer, respectively, as a function of the refractive index contrast $\Delta n$ and for some selected value of $n_2$. In both structures the surface field enhancement increases with $\Delta n$ as a consequence of the smaller mode length. At the same time, it decreases as $n_2$ increases, in particular when light is confined uniquely by TIR. The analysis of the field distribution (not shown here) reveals that in both cases this behaviour is caused by a shift of the maximum of the field far from the surface as the average refractive index of the structure increases. Finally, we notice that in Fig. \ref{fig5} (a) and (b), the largest achievable values of the surface fields are comparable. 
% FIGURE 6 Surface field ratio	
	\begin{figure}
		\centering
		\includegraphics[width=0.51\textwidth]{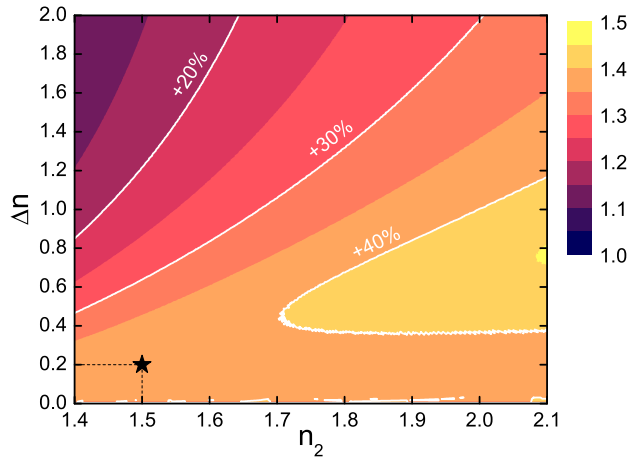}
		\caption{Contour plot of the ratio between the optimized surface fields of the BSW supported by the truncated periodic multilayer and the TE mode supported by the optimal asymmetric slab waveguide as a function of the lower index $n_2$ and the refractive index contrast $\Delta_n$. The star represents the refractive indices of the structures supporting the modes in Fig. 2. }
		\label{fig6}
	\end{figure}

For a complete comparison of the two structures, in Fig. 6 we plot the ratio between the largest surface field achievable for the TE BSW in the truncated periodic multilayer and that obtained for the fundamental TE mode in the asymmetric slab waveguide as a function of $n_2$ and $\Delta n$. The plot shows that the largest value of the surface field is always obtained for a BSW, with an enhancement up to 45\% with respect to the best possible scenario for an asymmetric slab waveguide with $n_2>1.7$ and $0.4<\Delta n<0.8$. Interestingly, this situation corresponds to the one in which the two structure have similar mode length, showing that the advantage of BSW is not due to a smaller mode length, but rather to the particular energy distribution in the structure associated with a confinement mechanism based on interference rather than TIR.

These results suggest that, in general, the need for a large surface field may not be sufficient to justify the choice of working with BSWs. Indeed, while truncated dielectric periodic multilayers can be characterized by a large electric field at the structure surface, the enhancement with respect to the best waveguide is usually quite modest, especially when $\Delta n$ increases. Naturally, the two structures have many other different properties which are not discussed in this work, but that could help identify the best solutions for specific applications.

% CONCLUSIONS
\section{Conclusions}
We carried out a thorough comparison between the fundamental TE mode supported by an asymmetric slab waveguide and the BSW supported by a truncated periodic multilayer in terms of two figures of merit, namely the mode length and the value of the electric field at the structure surface. We considered dielectric structures composed of materials having refractive indices varying between  $[1.4,3.4]$ with a refractive index contrast from 0 to 2, in the case in which absorption losses can be neglected.

Our results indicate that - perhaps surprisingly - modes confined by TIR are not always characterized by the smallest mode length (i.e. the largest electromagnetic energy density in a point). In particular, we found that for a sufficiently strong refractive index contrast ($\Delta n>0.6$) of the constituent materials, BSWs can exhibit smaller mode length than any mode supported by asymmetric slabs. This result is obtained for a proper truncation of the multilayer and the optimization of the unit cell according to a properly generalized $\lambda/4$-condition, typically used for the design of DBRs. 

In terms of the surface electric field, the overall trend indicates that BSWs always exhibit the largest surface field with respect to that of the best scenario in asymmetric slab modes. Here, we clarify that the enhancement with respect to the waveguide is limited to $45\%$, suggesting that in some cases simple slab waveguides might be indeed an equally valuable solution to achieve strong light-matter interaction at the structure surface. 

These results are wavelength independent and valid for a large class of multilayered structures made of semiconductors, oxides, and organic materials. Thus, we believe they will be of great help in the choice of the best platform to study and exploit the light-matter interaction in planar structures. 

\begin{acknowledgments}
The authors are grateful to John E. Sipe for stimulating discussions on light confinement in dielectric systems. 
\end{acknowledgments}
\appendix

% APPENDIX
\section{Appendix}
	\renewcommand\theequation{A\arabic{equation}}
	\setcounter{equation}{0}
	
	% APPENDIX - WG Results
	\subsection{Results for the asymmetric slab waveguide}
		The $z$-dependence of the fundamental TE mode in an asymmetric slab waveguide can be taken as 
\begin{align}
	E_m(z) = 
		\begin{cases}
			Ce^{-qz} 									\qquad &\textrm{if } z \geq 0\\
			C \left( \cos(hz) - \frac{q}{h}\sin(hz) \right) 		\qquad &\textrm{if } -d \leq z < 0\label{Scaled_Electric_field_definition}\\
			C \left( \cos(hd) + \frac{q}{h} \right) e^{p(z+d)} 		\qquad &\textrm{if } z < -d
		\end{cases}
\end{align}\label{electric_field}
where $C$ is a normalization constant, $h$, $q$ and $p$ are the transversal components of the wavevector in each layer, i.e.
\begin{align}
	&h = \sqrt{\left(\frac{\omega}{c}n_{1}\right)^2 - \beta^2}\\
	&q = \sqrt{\beta^2 - \left(\frac{\omega}{c}n_{e}\right)^2}\\
	&p = \sqrt{\beta^2 - \left(\frac{\omega}{c}n_{2}\right)^2},
\end{align}
and $n_1$, $n_2$ and $n_e$ are respectively the refractive indices of the core layer, of the substrate and of the external medium, as in figure (\ref{fig1}).

The normalization constant $C$ can be evaluated by normalizing the field, that is by imposing that each mode transports the energy of a photon
\begin{equation}
	A\int_{-\infty}^{+\infty} \varepsilon(z) |E(z)|^2 dz = \frac{\hbar \omega}{2},
\end{equation}
where $A$ is a normalization area (taken as $A = 1\ m^2$ throughout this article). This translates to 
\begin{equation}
	C = \sqrt{\frac{\hbar \omega}{2\left(I_{ext} + I_{core} + I_{sub}\right)}},
\end{equation}
where
\begin{subequations}\label{I_definition}
	\begin{align}
		I_{ext}   &= \frac{C^2\varepsilon_{ext}}{2q} \\
		I_{sub}  &= \frac{C^2 \varepsilon_{sub}\left[ \cos(hd) + \frac{q}{h}\sin(hd) \right]^2}{2p} \\
		I_{core} &= \frac{C^2 \varepsilon_{core}}{2} \bigg[ d + \frac{\sin(2hd)}{2h} + \frac{q^2}{h^2}\bigg(d - \frac{\sin(2hd)}{2h}\bigg) + \nonumber \\
			     &\qquad \qquad +\frac{q}{h^2}\bigg(1 - \cos(2hd)\bigg)\bigg] 
	\end{align}
\end{subequations}

In order for the slab to support the $m$-th TE mode, its thickness must be capped by
\begin{equation}
	\frac{d_{max}}{\lambda} = \frac{1}{2\pi\sqrt{n_{core}^2 - n_{sub}^2}} \left[ m\pi + \arctan\sqrt{\frac{n_{sub}^2 - n_{ext}^2}{n_{core}^2 - n_{sub}^2}} \right].\label{d_over_lambda}
\end{equation}

Here we have focussed on the fundamental TE0 mode, since we were interested in maximum confinement, therefore in our simulations we set $m = 0$.

The maximum of the field can be calculated analytically by deriving the expression given in eq. (\ref{electric_field}); the result is
\begin{equation}
	E_{max} = C\frac{\sqrt{h^2 + q^2}}{h}
\end{equation}

When the electric field is defined as above, the surface field is simply $E_m(0) = C$, thus it can be calculated by properly normalizing the electric field.

	% APPENDIX - WG Results - Scalability
%	\subsubsection{Scalability}
%	All the figures defined so far can be shown to depend essentially on $x = \beta d$ and $y = d/\lambda_0$; for example, 
%	\begin{align}
%		&hd = \sqrt{\left(2 \pi n_{core} y\right)^2 - x^2}\\
%		&qd = \sqrt{x^2 - \left(2 \pi n_{ext} y\right)^2}\\
%		&pd = \sqrt{x^2 - \left(2 \pi n_{sub} y\right)^2},
%	\end{align}
%	and
%	\begin{equation}
%		E_{max} = C\sqrt{1 + \left(\frac{qd}{hd}\right)^2}.	
%	\end{equation}
%	We can then define two new figures, namely $\tilde{S}_{field}$ and $\tilde{V}_{mod}$, as
%	\begin{align}
%		\tilde{S}_{field} &= C\lambda = \frac{h_{Planck} c}{2y(\tilde{I}_{ext} + \tilde{I}_{core} + \tilde{I}_{sub})} \label{Scaled_SField}\\
%		\tilde{V}_{mod} &= \frac{L_{mod}}{\lambda} = \frac{L_{mod}}{d} \cdot \frac{d}{\lambda} = \nonumber \\
%				         &= \frac{1}{\varepsilon_{core}|E_{max}|^2} \left( \tilde{I}_{ext} + \tilde{I}_{core} + \tilde{I}_{sub}) \right) \label{Scaled_Vmod}
%	\end{align}
%	where $\tilde{I}_j = I_j/d$, as defined in eq. (\ref{I_definition}).
%
%	Using $x$ and $y$ as variables, the light lines become
%	\begin{equation}
%		y_j = \frac{1}{2\pi n_j} x.\label{Scaled_LL}
%	\end{equation}

	% APPENDIX - BSW
	\subsection{Results for truncated periodic multilayer}
	Let us now consider a periodic multilayer with a unit cell consisting of refractive indices $n_{1}$ and $n_{2}$ and thicknesses $d_1$ and $d_2$, as shown in fig. (\ref{fig1}). Let the period of this structure be $\Lambda = d_1 + d_2$.

	The electric field in each layer can be written in terms of forward and backward components, \emph{i.e.}
	\begin{equation}
		E_j(z) = E_j^+ e^{i k_j z} + E_j^- e^{-i k_j z}
	\end{equation}
	Components across layers are linked via the interface matrices, \emph{i.e.}
	\begin{equation}
\left( \begin{array}{c}
E_j^+ \\
E_j^-
\end{array}  \right) =
\frac{1}{2t_{j,j-1}}
\left( \begin{array}{cc}
1 & r_{j,j-1} \\
r_{j,j-1} & 1
\end{array}  \right)
\left( \begin{array}{c}
E_{j-1}^+ \\
E_{j-1}^-
\end{array}  \right),
	\end{equation}

where the $r$ and $t$ Fresnel coefficients for TE polarization are defined as
\begin{align}
	r_{j,j-1} = \frac{w_j - w_{j-1}}{w_j + w_{j-1}}\\
	t_{j,j-1} = \frac{2w_j}{w_j + w_{j-1}};
\end{align}
the propagation of the field components within a layer can be obtained by resorting to \emph{propagation matrices}, \emph{i.e.}
	\begin{equation}
\left( \begin{array}{c}
E_j^+(z + d_j) \\
E_j^-(z + d_j)
\end{array}  \right) =
\left( \begin{array}{cc}
e^{i w_j d_j} & 0 \\
0 & e^{-i w_j d_j}
\end{array}  \right)
\left( \begin{array}{c}
E_{j-1}^+(z) \\
E_{j-1}^-(z)
\end{array}  \right),
	\end{equation}

Carrying out products of \emph{interface} and \emph{propagation} matrices allows one to calculate the \emph{transfer matrix} for the photonic system under scrutiny, and ultimately to express field components in each layer of the structure.
Once these terms are known, it is possible to calculate the FoMs for the BSW.

% Calculation of \sigma
In order for the multilayer to support a BSW at a given $\beta$, the first layer must be truncated to a length $\sigma d_1$, as explained in \cite{liscidini_JOSAB}, where the truncation factor $0 < \sigma < 1$ is given by
\begin{equation}\label{sigma_TE}
	\sigma = \frac{1}{2iw_1L_1}\log\left[\frac{M_{12}(q_{ext} - iw_1)}{(M_{11} - e^{-q\Lambda})(iw_1 + q_{ext})}\right],
\end{equation}
where $q_{ext} = \Im{\{w_{ext}\}}$.

% External field contribution to L_{mod}
The evanescent field in the semi-infinite external medium contributes to the overall mode length $L_{mod}$ with
	\begin{equation}\label{V_mod_ext}
		L_{mod}^{ext} = \varepsilon_{ext} \frac{|E_{ext}^-|^2}{2q_{ext}}.
	\end{equation}

% The contribution due to the j-th layer
The contribution to the mode length due to each layer can be calculated according to the general definition given in eq. (\ref{Lmod}), where the integral is extended only to the layer under scrutiny. The resulting expression is
	\begin{equation}\label{V_mod_each_layer}
		L_{mod}^j = \varepsilon_j [|E_j^+|^2 + |E_j^-|^2] + \varepsilon_j \Im \left [\frac{E_j^+ E_j^{-, *}(e^{2iw_jd_j} - 1)}{w_j}\right],
	\end{equation}
where $w_j = \sqrt{(2\pi/\lambda_0)^2\varepsilon_j - \beta^2}$ is the transversal component of the wave vector in each layer.

% The geometric factor
	In order not to include finite-size effects, we have considered only semi-infinite structures, \emph{i.e.} multilayers consisting of an infinite repetition of a bilayer unit cell surmounted by a truncated layer. According to Bloch's theorem, $E(z + \Lambda) = e^{i k_{Bloch} z }E(z)$, and since BSWs live in the PBG, the Bloch wave vector $k_{Bloch}$ is imaginary and $E(z + \Lambda) = e^{-q\Lambda}E(z)$. This means that after one period the field has decayed by $e^{-q\Lambda}$, and therefore its intensity has decreased by the square of this expression. To calculate the integral of the modulus square of the electric field due to the whole semi-infinite periodic structure, one can then calculate $I_{u.c.}$, \emph{i.e.} the integral extended to the first unit cell, multiply it by the sum of the geometric series of ratio $k = e^{-2q\Lambda}$, and sum the contribution due to the truncation layer and the external medium, so that
	\begin{equation}
		L_{mod}^{semi-\infty} = K\left[V_{ext} + V_{trunc} + \frac{I_{u.c.}}{1 - e^{-2q\Lambda}} \right],
	\end{equation}
where $K$ is the inverse maximum electromagnetic energy density appearing in eq. (\ref{Lmod}), and $I_{u.c.}$ is given by
\begin{align}
	I_{u.c.} = &\varepsilon_1 [|E_1^+|^2 + |E_1^-|^2] + \varepsilon_1 \Im \left [\frac{E_1^+ E_1^{-, *}(e^{2iw_1 d_1} - 1)}{w_1}\right] \\
		       +&\varepsilon_2 [|E_2^+|^2 + |E_2^-|^2] + \varepsilon_2 \Im \left [\frac{E_2^+ E_2^{-, *}(e^{2iw_2 d_2} - 1)}{w_2}\right].
\end{align}

% Maximum field intensity
	To complete the expression for the mode length, we need to find an expression for the maximum field intensity contained in this prefactor $K$; this can be obtained by deriving the general term in eq. (\ref{V_mod_each_layer}), which yields
	\begin{equation}
		\frac{\partial L_{mod}^j}{\partial z} = 0 \to \Im\left[E_j^+E_j^{-,*}e^{2ik_jz}\right] = 0.
	\end{equation}
	The solution we are after is then
	\begin{equation}
		z_{max} = -\frac{1}{2k_j}\arctan\left\{\phi(E_j^+E_j^{-,*})\right\},
	\end{equation}
	where $\phi(E_j^+E_j^{-,*})$ is the phase of the complex number $E_j^+E_j^{-,*}$, \emph{i.e.}
	\begin{equation}
		\phi(E_j^+E_j^{-,*}) = \frac{\Im[E_j^+E_j^{-,*}]}{\Re[E_j^+E_j^{-,*}]}.
	\end{equation}
	To calculate the maximum field intensity, it is sufficient to calculate $\varepsilon_j E_j(z_{max})$ in the truncation layer and the unit cell, and to select the maximum value among them. 
	\begin{equation}\label{max_field_square}
		(\varepsilon|E|^2)_{max} = \max\limits_j \left\{ \varepsilon_j |E(z_{max, j})|^2 \right\}
	\end{equation}

% BSW mode length
	By combining all the results, the overall mode length is the sum of the terms given by eq. (\ref{V_mod_each_layer}) and the external contribution (\ref{V_mod_ext}), divided by the maximum field intensity given by eq. (\ref{max_field_square}):
	\begin{equation}\label{BSW_VMod}
		L_{mod} = \frac{1}{\max\limits_j \left\{ \varepsilon_j |E(z_{max, j})|^2 \right\}} \left[ V_{ext} + V_{trunc} + \frac{I_{u.c.}}{1 - e^{-2q\Lambda}} \right]
	\end{equation}

% BSW Surface Field
	If we assume that the $x$ axis lies on the interface between the truncation layer and the semi-infinite external dielectric medium, the surface field is given essentially by
	\begin{equation}\label{BSW_SField}
		S_{field} = E_{ext}^-.
	\end{equation}

% Bibliography 

%\bibliographystyle{te}
\bibliography{Article}

\end{document}